\documentclass[floats,aps,twocolumn,showpacs]{revtex4}

\usepackage{graphicx}
\usepackage{color}
\usepackage{amsmath}

\begin{document}

\newcommand{\bk}{{\bf k}}
\newcommand{\BSCCO}{{Bi$_2$Sr$_2$CaCu$_2$O$_{8+x}$}}
\newcommand{\YBCO}{{YBa$_2$Cu$_3$O$_{7-\delta}$}}
\def\k{{\bf k}}
\def\q{{\bf q}}

\title{Thermodynamic transitions in inhomogeneous $d$-wave superconductors}

\author{ Brian M. Andersen$^{1,2}$, Ashot Melikyan$^1$, Tamara S. Nunner$^3$,
 and  P. J. Hirschfeld$^{1,4}$}

\affiliation{$^1$Department of Physics, University of Florida,
Gainesville, Florida 32611-8440, USA\\ $^2$Laboratoire de Physique
Quantique (CNRS), ESPCI, 10 Rue de Vauquelin, 75231 Paris, France\\
$^3$Institut f\"{u}r Theoretische Physik, Freie Universtit\"{a}t
Berlin, Arnimallee 14, 14195 Berlin, Germany\\
$^4$Laboratoire de Physique des Solides, Universit\'{e} Paris-Sud,
91405 Orsay, France}

\date{\today}

\begin{abstract}
We study the spectral and thermodynamic properties of
inhomogeneous $d$-wave superconductors within a model where the
inhomogeneity originates from atomic scale pair disorder. This
assumption has been shown to be consistent with the small charge
and large gap modulations observed by scanning tunnelling
spectroscopy (STS) on \BSCCO.  Here we calculate the specific heat
within the same model, and show that it gives a semi-quantitative
description of the transition width in this material. This model
therefore provides a consistent picture of both surface sensitive
spectroscopy and bulk thermodynamic properties.

\end{abstract}


\maketitle

{\it Introduction.} There is accumulating evidence from STS that
at least some of the families of superconducting cuprate materials
are "intrinsically" inhomogeneous at the
nanoscale\cite{cren,davisinhom1,Kapitulnik1,davisinhom2}, in the
sense that all samples of the given material exhibit electronic
disorder over length scales of $\sim$ 25\AA~ which cannot be
removed by any standard annealing procedure.  In particular,
differential conductance maps of \BSCCO (BSCCO) reveal the
existence of spectral peaks reminiscent of the coherence peaks of
a homogeneous $d$-wave superconductor whose energy varies by a
factor of 2-3 over the sample. There is currently widespread
interest in this phenomenon, but no consensus as to its origin,
nor even as to whether the gaps between these peaks may be taken
to correspond directly to the local superconducting order
parameter in the sample, or whether some are related to a second,
competing order\cite{KivelsonRMP}. While these experiments have
been proposed as evidence of a general nanoscale inhomogeneity in
cuprates, there have been several criticisms. One class of
objections points out that other materials, particularly \YBCO
(YBCO) exhibit much narrower NMR linewidths, suggesting that in
this material at least, the electronic inhomogeneity cannot be a
pronounced property of the bulk\cite{Bobroffcritique}. A stronger
critique has been offered by Loram {\sl et al.}\cite{Loram} even
with regard to the BSCCO material. These authors point out that if
one associates with each "gap patch" a local doping corresponding
to the hole density required to produce a gap of that size in the
average phase diagram, then the distribution of the doping would
be of order $\Delta p\sim$0.1. Such a large range of local doping
levels would in turn produce a large inhomogeneous distribution of
regions with different $T_c$'s, sufficient to yield a transition
width $\Delta T_c$ of order the average $T_c$ itself. Since this
is contrary to specific heat measurements\cite{junod99,Loram2},
these authors conclude that the inhomogeneity is either a surface
phenomenon, or represents a distribution of antinodal scattering
lifetimes rather than of order parameters. Clearly it is very
important to determine whether results from surface sensitive
probes reflect the bulk properties of these materials.

Recently, localized resonances were imaged by STS at a bias of
-960 meV and identified as the O interstitials primarily
responsible for doping the BSCCO material\cite{DavisScience05}.
This work also noted the  strong positive correlation between the
positions of these dopants and the local spectral gap, and
furthermore argued that the charge inhomogeneity in the sample was
considerably smaller than anticipated on the basis of previous
scanning tunnelling topographs integrated to smaller bias
voltages. Nunner {\sl et al.}\cite{NAMH05} then argued that these
and a number of other experimental observations\cite{nunner2}
could be explained by the simple hypothesis that the disorder
caused by the dopant atoms was primarily in the Cooper rather than
density channel, i.e. that each dopant modulated the BCS pair
interaction on an atomic scale.

In this Letter we argue that the lack of strong charge modulations
within the pair disorder model of Ref. \onlinecite{NAMH05} also
allows one to avoid the argument of Loram {\sl et al.}\cite{Loram}
regarding the transition width.  The model, with parameters chosen
to reproduce the gap maps and other correlations of the STS data
on BSCCO, is shown to yield results within mean field theory which
are  consistent with a relative specific heat transition width,
$\Delta T_c/T_c$, of $\sim 20\%$, despite the fact that at $T=0$
gap modulations of order 100\% are observed. The observed widths
must therefore be attributable to pair disorder alone since the
samples are homogeneous at the {\it mesoscale} (as determined,
e.g., by optical and scanning electron microscopy). Predictions of
the theory for the formation and persistence of superconducting
islands near the transition should be verifiable by
high-temperature STS measurements.

{\it Model.}  There are many studies of inhomogeneous
superconductivity\cite{BalatskyRMP}, only few of which are
directly relevant to the questions addressed below.  If one adds
nonmagnetic disorder to an ordinary $s$-wave superconductor, one
expects only small changes in superconducting one-particle
properties such as the gap and $T_c$ due to Anderson's
argument\cite{Andersontheorem}. Nevertheless, as disorder
increases and the mean free path $\ell$ becomes comparable to the
Fermi wavelength $\lambda_F$, localization effects can destroy
superconductivity. Recently, local aspects of this transition were
studied by Ghosal {\sl et al.} \cite{Ghosaletal}, who showed using
numerical solutions of the Bogoliubov-de Gennes (BdG) equations
including nonmagnetic disorder that near the transition the highly
disordered system separates into islands with finite order
parameter surrounded by an insulating sea. In the $d$-wave case,
where ordinary disorder is pair breaking and the transition occurs
when $\ell\sim\xi_0$, where $\xi_0$ is the coherence length,
several numerical studies have investigated local properties of
disordered
systems\cite{Franzetal,Atkinson1,Ghosaletaldwave,Atkinson2}. Only
Ref. \onlinecite{Franzetal} considered the finite temperature
transition, however, pointing out that the inhomogeneity induced
by ordinary disorder (in this case the suppression of the order
parameter around strong scattering centers) strongly affects the
transition, leading to a $T_c$ onset (defined by the formation of
superconducting islands as temperature $T$ is reduced from above),
considerably higher than the $T_c$ predicted by the usual
disorder-averaged theory of $T_c$ suppression, analogous to the
theory of Abrikosov and Gorkov\cite{AG}. The width of the
transition was not discussed explicitly in this work, nor was the
effect of pairing disorder considered.

More recently, it has been proposed that $T_c$ can even be increased
by local inhomogeneities compared to its value for the homogeneous
system\cite{podolsky,aryanpour}. Within a weak coupling BCS
framework it was shown that periodic modulations of the pairing
and/or electron density can lead to a substantial enhancement of
$T_c$ when the characteristic modulation length scale is of order
$\xi_0$\cite{podolsky}. This agrees with recent numerical
calculations for the attractive Hubbard model with various disorder
distributions of the interaction, verifying that $T_c$ enhancement
by inhomogeneity is a general result arising from the proximity
effect\cite{aryanpour}. These calculations did not investigate the
details of the transition below the $T_c$ where the local order
parameter $\Delta$ first nucleates.

In the following model calculation we use the standard $d$-wave BCS
Hamiltonian
\begin{equation}
\label{eq:hamiltonian} \hat{H}\!=\!\sum_{{\langle ij
\rangle}\sigma}\! t_{ij} \hat{c}_{i\sigma}^\dagger
\hat{c}_{j\sigma} +\! \sum_{i\sigma} \!
(V_i-\mu)\hat{c}_{i\sigma}^\dagger \hat{c}_{i\sigma} \!+\!
\sum_{\langle ij \rangle} \! \left( \Delta_{ij}
\hat{c}_{i\uparrow}^\dagger \hat{c}_{j\downarrow}^\dagger \!+\!
\mbox{H.c.} \! \right)\!,\!
\end{equation}
where in the first term we include nearest $t$ and next-nearest
$t^\prime=-0.3t$ neighbor hopping. For the chemical potential
$\mu$, we set $\mu=-t$ in order to model the Fermi surface of
BSCCO near optimal doping $\sim 16 \%$. $\sum_{\langle ij
\rangle}$ denotes summation over neighboring lattice sites $i$ and
$j$. Disorder is included by the impurity potential $V_i =V_0 f_i$
where $f_i=\sum_s \exp(-r_{is}/\lambda)/r_{is}$, $r_{is}$ being
the distance from a defect $s$ to the lattice site $i$ in the
plane. Distances are measured in units of $\sqrt{2}a$, where $a$
is the Cu-Cu distance. Note that the particular Yukawa form of
$f_i$ is merely a convenient way to vary the smoothness of the
potential landscape through the parameter $\lambda$. The $d$-wave
order parameter $\Delta_{ij}=g_{ij} \langle \hat{c}_{i\uparrow}
\hat{c}_{j\downarrow} - \hat{c}_{j\downarrow}
\hat{c}_{i\uparrow}\rangle/2$ is determined self-consistently by
iterations of
\begin{equation}
\Delta_{ij}=\frac{g_{ij}}{2} \sum_n \left( u_n(i) v_n(j) + v_n(i)
u_n(j)\right) \tanh(\frac{E_n}{2 T}),
\end{equation}
until convergence is achieved. Here, $\{ E_n , u_n , v_n\}$ is the
eigensystem resulting from diagonalization of the BdG equations
associated with Eq.(\ref{eq:hamiltonian}). The pairing interaction
$g_{ij}$ varies in space relative to its homogeneous value
$g_0=1.16t$ as $g_{ij}=g_0+\delta g (f_i+f_j)/2$, where $\delta g$
is the modulation amplitude and $i,j$ are nearest neighbors. In
the following, when including potential ($\tau_3$ channel,
$V_0\neq 0$) or pair ($\tau_1$ channel, $\delta g\neq 0$)
disorder, we make sure to adjust $\mu$ and $g_0$, respectively, in
order to maintain the same doping and average gap as the
corresponding homogeneous system. Note that in this approach the
inclusion of spatial pair potential variations is purely
phenomenological. However, a likely candidate for the pairing
modulations is the oxygen dopants. Various possible microscopic
origins for this unusual disordered state have been discussed in
Refs. \onlinecite{NAMH05,nunner2}. There it was also argued that
$\tau_3$ disorder alone cannot account for the STM data.
\begin{figure}[b]
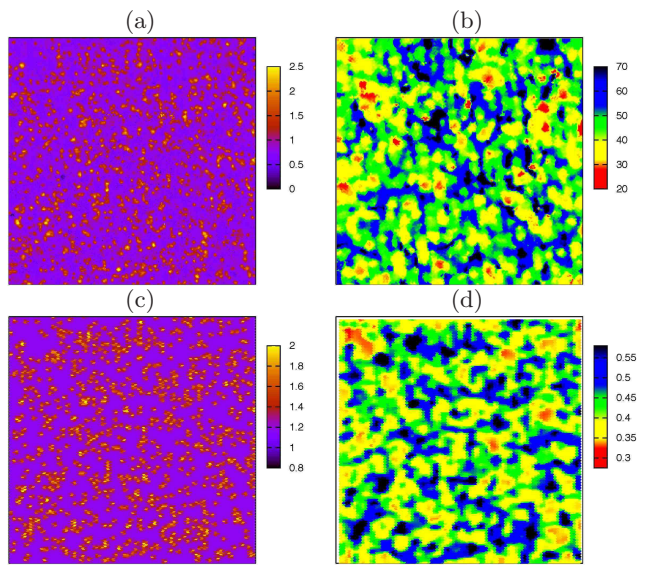

\begin{minipage}{.49\columnwidth}
\hspace{-0.5cm}(a)\\
\includegraphics[clip=true,bb=242 148 650
485,width=.95\columnwidth]{mapdiluted_lowres.epsf}
\end{minipage}
\begin{minipage}{.49\columnwidth}
\hspace{-0.5cm}(b)\\
\includegraphics[clip=true,bb=242 148 650
485,width=.95\columnwidth]{41026A00GapMap_gp.epsf}
\end{minipage}\\
\begin{minipage}{.49\columnwidth}
\hspace{-0.5cm}(c)\\
\includegraphics[clip=true,bb=242 148 650
485,width=.95\columnwidth]{newImpPot90Vi0001lam05rz05.epsf}
\end{minipage}
\begin{minipage}{.49\columnwidth}
\hspace{-0.5cm}(d)\\
\includegraphics[clip=true,bb=242 148 650
485,width=.95\columnwidth]{GapMap_lowres.epsf}
\end{minipage}
\caption{(Color online.) (a) Experimental $dI/dV$ map [arb. units]
at -960 meV of an optimally doped BSCCO sample from Ref.
\onlinecite{DavisScience05}. (b) Experimental gap map [meV] in the
same region as (a). (c) The theoretical impurity potential
extracted from (a) assuming distance from plane $r_z=0.5$ and
$\lambda=0.5$. (d) The gap map resulting from using (c) as the
'pairing potential' in the BdG equations with $\delta g=1.5t$.
Both (c) and (d) are shown in units of $t$.} \label{fig:2Dmaps}
\end{figure}
%
%
%

{\it Results.} Fig. \ref{fig:2Dmaps} shows, in a field of view
(FOV) of $49\mbox{nm}\times 49\mbox{nm}$, the experimental LDOS at
-960 meV (a), and the corresponding gap map (b) obtained by
McElroy {\sl et al.}\cite{DavisScience05}. The bright resonances
in \ref{fig:2Dmaps}(a) reveal the location of the oxygen dopants.
In Fig. \ref{fig:2Dmaps}(c) we show the impurity potential
generated by using each of these dopants (833 in the above FOV) as
a defect located out of the CuO$_2$ plane. This map resembles (a)
as it should. The experimental FOV is modeled as a $90\times 90$
lattice system rotated 45 degrees with respect to  the 3.83 \AA
$~$ long Cu-Cu bonds, i.e. it includes $2\times 90\times 90$ sites
and is aligned with the experimental FOV\cite{comment1}. The
theoretical gap map (as extracted from the LDOS) using (c) as a
$\tau_1$ potential with $\delta g=1.5t$ at $T=0$ is shown Fig.
\ref{fig:2Dmaps}(d). We find that  gap maps consistent with
experiment are found for $1.0<\delta g/t<2.0$. The correlation
coefficient between the gap maps \ref{fig:2Dmaps}(b) and
\ref{fig:2Dmaps}(d) is 0.17\cite{comment3}, a reasonably large
value given the simplicity of the pure $\tau_1$ calculation and
the fact that the correlation between the dopant positions and the
experimental gap map is 0.3\cite{DavisScience05}.
\begin{figure}[b]
\begin{minipage}{.49\columnwidth}
(a)\\[-.3cm]
\includegraphics[clip=true,bb=120 200 500 660,width=.90\columnwidth,angle=270]{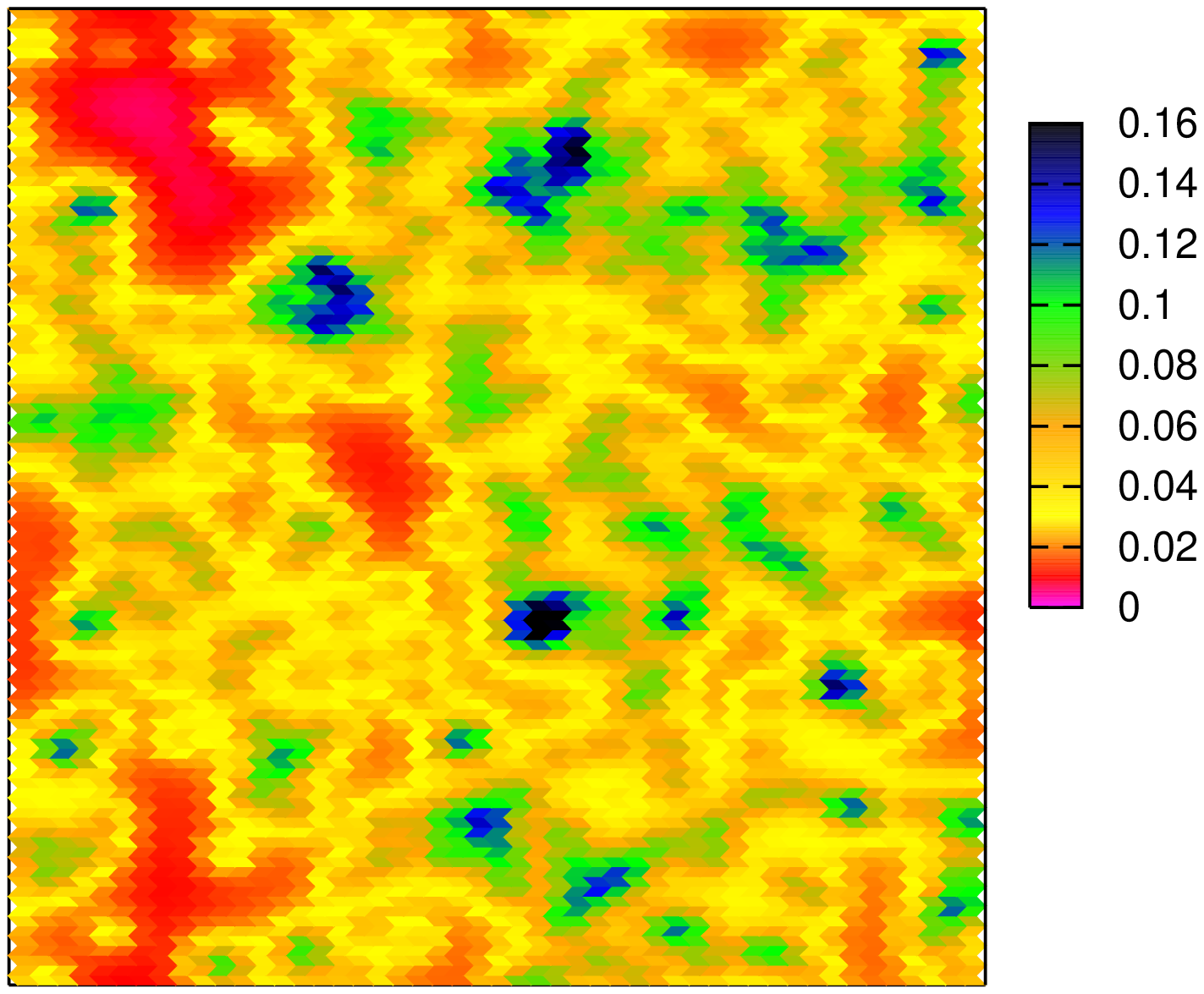}
\end{minipage}
\begin{minipage}{.49\columnwidth}
(b)\\[-.3cm]
\includegraphics[clip=true,bb=120 200 500 660,width=.90\columnwidth,angle=270]{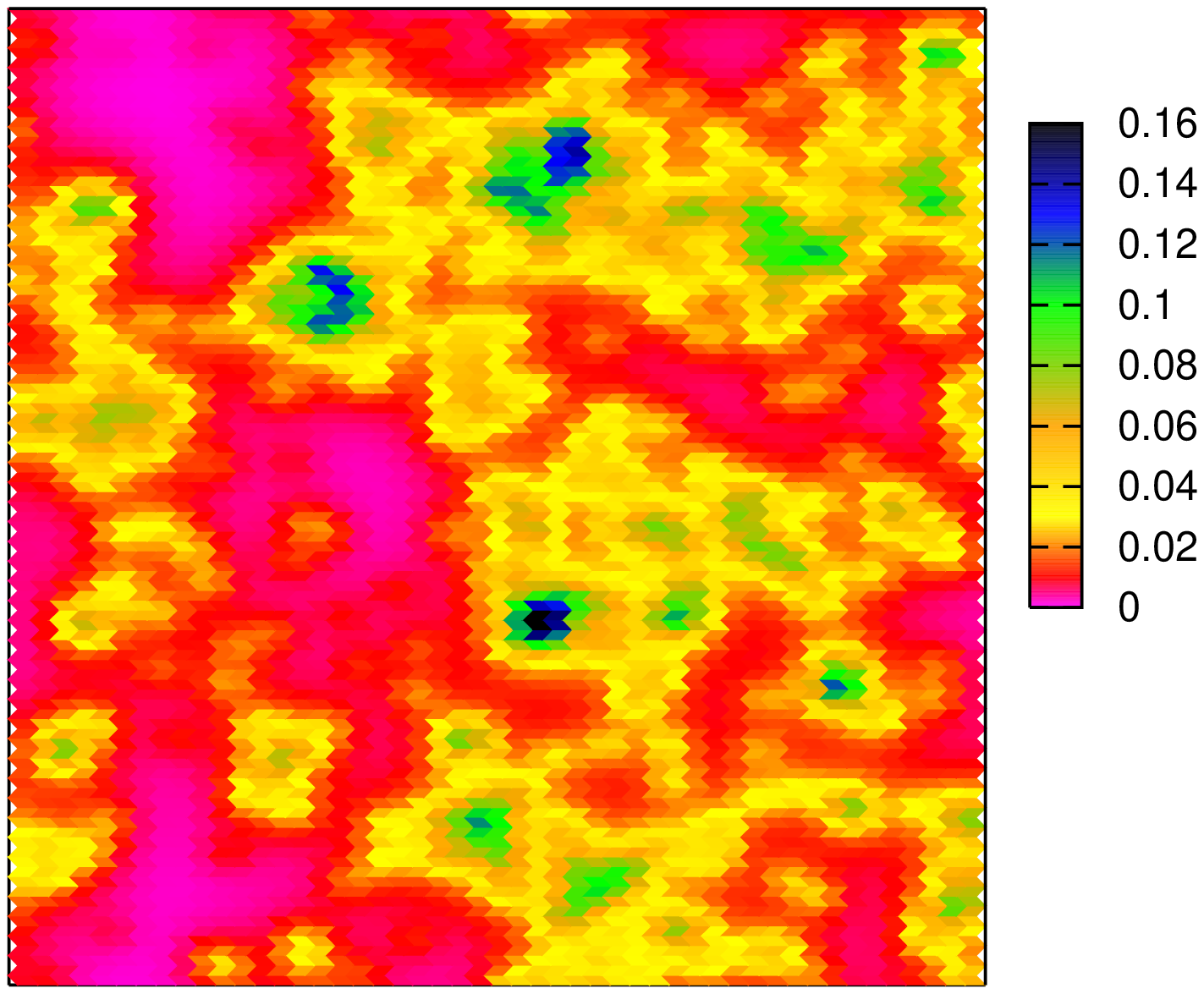}
\end{minipage}\\[-.1cm]
\begin{minipage}{.49\columnwidth}
(c)\\[-.3cm]
\includegraphics[clip=true,bb=120 200 500 660,width=.90\columnwidth,angle=270]{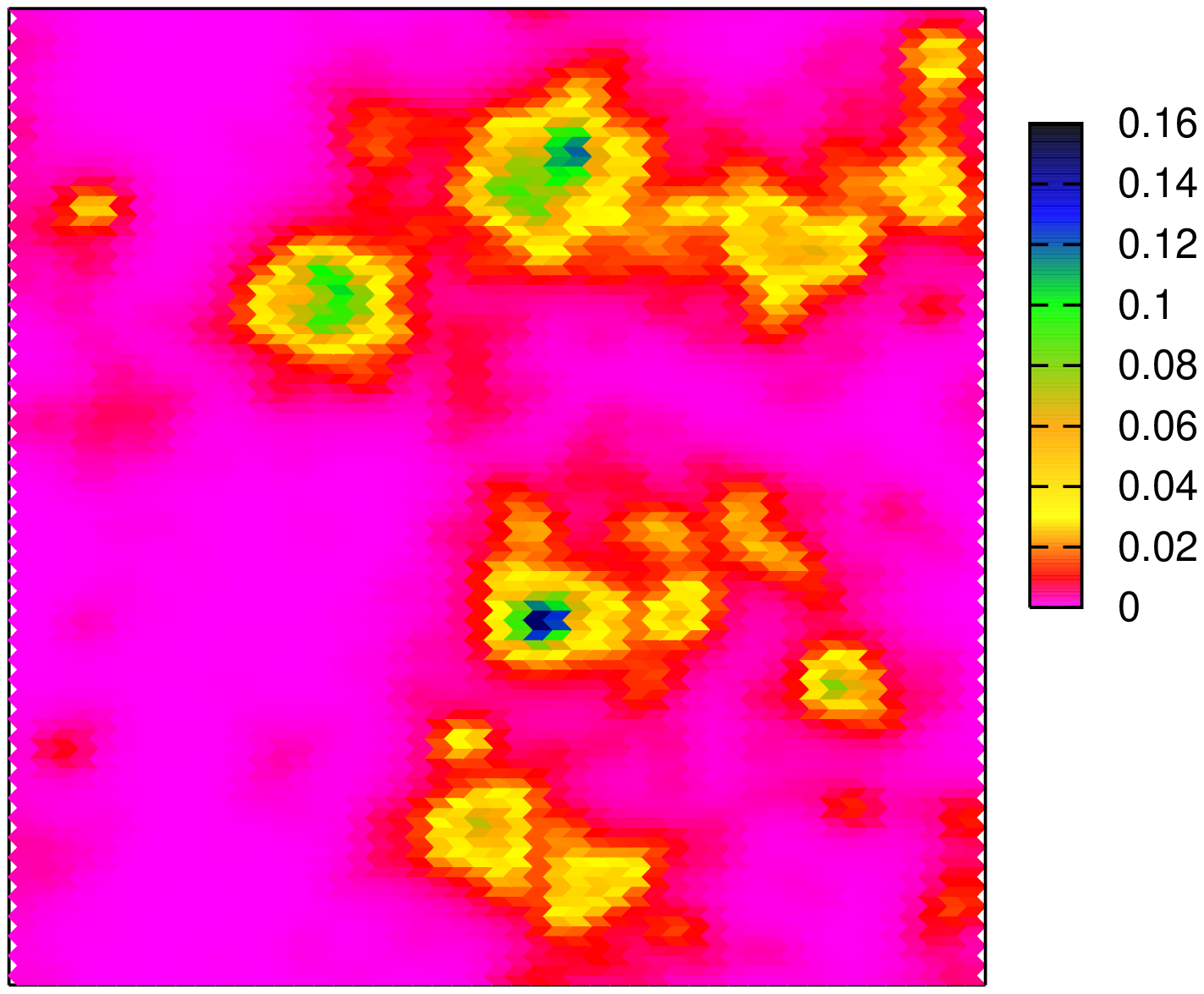}
\end{minipage}
\begin{minipage}{.49\columnwidth}
(d)\\[-.3cm]
\includegraphics[clip=true,bb=120 200 500 660,width=.90\columnwidth,angle=270]{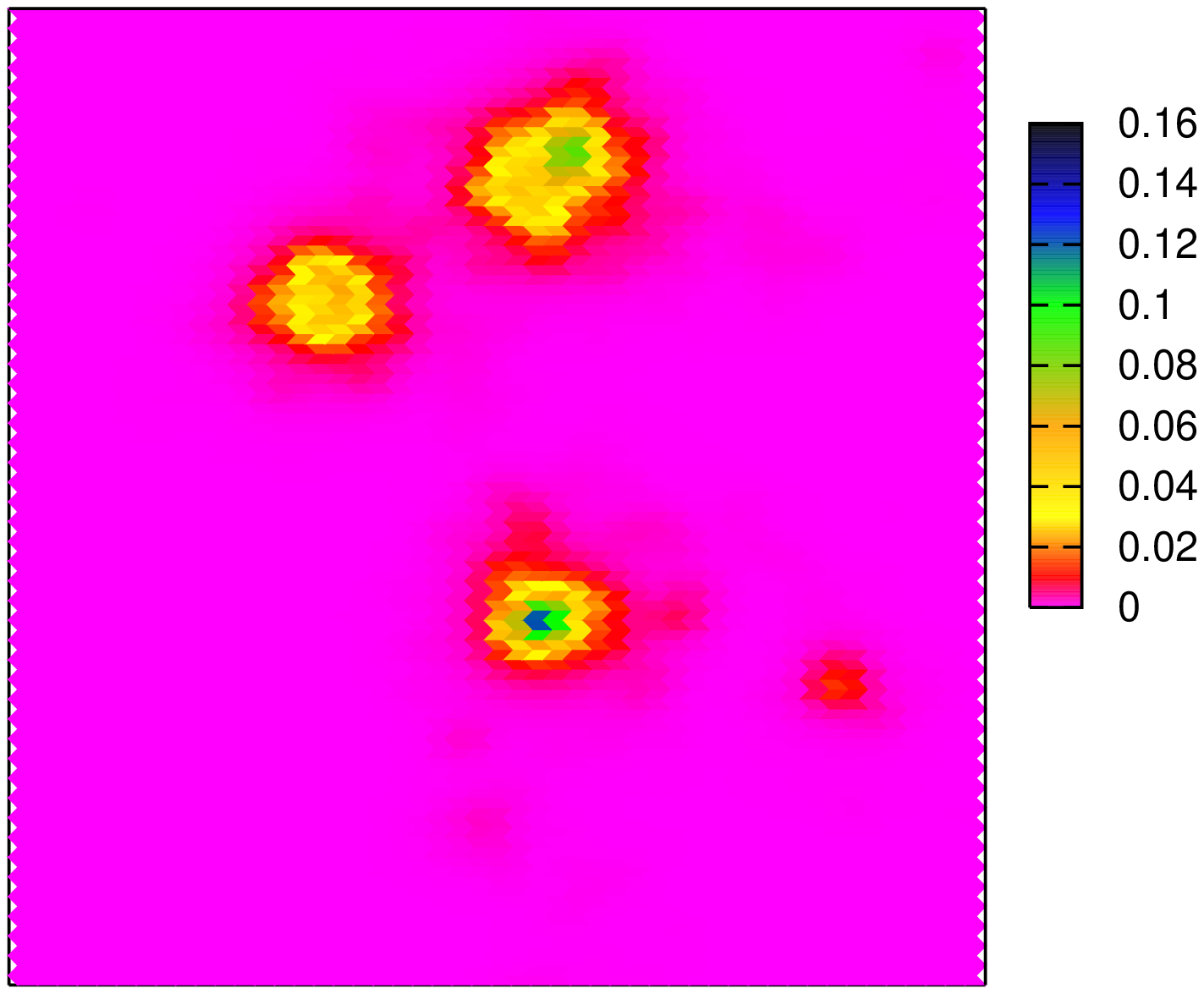}
\end{minipage}
\caption{(Color online) OP maps, parameters from Fig. 1(d):
$T\!=\!0.18t$ (a), $T\!=\!0.20t$ (b), $T\!=\!0.22t$ (c), and
$T\!=\!0.24t$ (d).} \label{fig:finiteTmaps}
\end{figure}

Now we turn to the discussion of the results for $T\neq 0$. Below,
we set $N=50$ (corresponding to the upper left 5/9th square of
each image in Fig.\ref{fig:2Dmaps}) and study the resulting order
parameter (OP) maps, entropy $S(T)$, and specific heat $C(T)$ with
focus on the temperature region near $T_c$. The number of
iterations necessary for converged results increases dramatically
near $T_c$, whereas just a few degrees away from $T_c$ we
typically find that 25-50 iterations suffice. In Fig.
\ref{fig:finiteTmaps}, we show the OP map
$\Delta_i=\left(\Delta_{i,i+\hat{x}}+\Delta_{i,i-\hat{x}}-\Delta_{i,i+\hat{y}}-
\Delta_{i,i-\hat{y}}\right)/4,$ for temperatures near $T_c$. The
bulk transition temperature for the homogeneous system is
$T_{c0}=0.182t$. Here, one clearly sees how separate islands of
finite $\Delta_i$ start forming above $T_{c0}$ when cooling down
and eventually overlap at lower $T$. In principle, STS
measurements at $T$ close to $T_c$ should be able to probe these
superconducting islands.

\begin{figure}[t]
\begin{center}
\includegraphics[width=0.95\columnwidth,height=5.0cm]{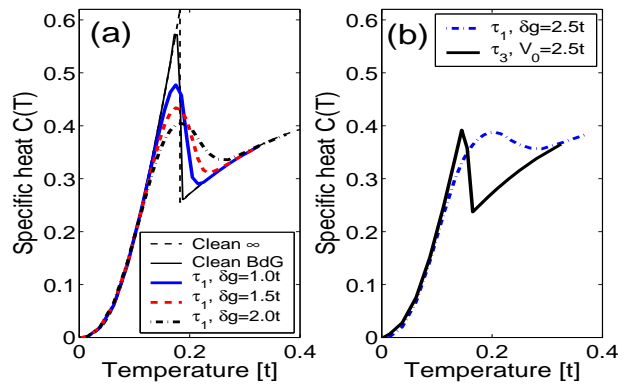}
\end{center}
\caption{(Color online) Specific heat $C(T)$ for the impurity
distribution of Fig. \ref{fig:2Dmaps}(c) vs $T$: (a) clean
$50\times 50$ BdG system (solid, black line), "infinite" clean
system (dashed, black line), and $\tau_1$ disordered with $\delta
g=1.0t$ (thick solid, blue line); $\delta g=1.5t$ (thick dashed,
red line); $\delta g=2.0t$ (thick dash-dotted, black line). (b)
Comparison of $C(T)$ for the Nambu channels $\tau_1$ and $\tau_3$
with otherwise identical  impurity parameters.}\label{fig:Entropy}
\end{figure}

In order to address the effect of the inhomogeneity on the
transition widths, we first calculate the quasiparticle entropy
$S(T)$
according to the well-known expression
\begin{equation}
S\!=\!-2\!\sum_{E_n>0} \!\left[ f(E_n) \ln f(E_n)\! +\! f(-E_n)
\ln(f(-E_n)) \right],
\end{equation}
where $f(E)$ is the Fermi distribution function. From the
resulting entropy $S(T)$ curve we extract the specific heat at
constant volume $C(T)$ by the usual expression $C(T)=T(\partial
S/\partial T)$. Fig. \ref{fig:Entropy}(a) compares the electronic
specific heat for the clean and pair disordered systems described
by different disorder strengths $\delta g$. In Fig.
\ref{fig:Entropy}(b) we compare $C(T)$  for the different Nambu
channels $\tau_1$ and $\tau_3$.
We can check the BdG results by comparing to the pure "infinite"
system result (black, dashed line in Fig. \ref{fig:Entropy}(a))
obtained by replacing $E_n\rightarrow E_\k\equiv
\sqrt{\epsilon_{\bk}^2+\Delta_{\bk}^2}$, with $\epsilon_{\bk}$ the
normal state dispersion with hopping integrals and chemical
potential identical to the values given above, and
$\Delta_{\bk}=\Delta_0(T) (\cos k_x - \cos k_y)/2$ with
$\Delta_0(T)$ solved selfconsistently from the usual gap equation.
It is clear that the $50\times 50$ BdG lattice problem is large
enough to capture the sharp transition width of the clean system.

Conventional potential disorder corresponding to $\sim 10\%$
dopant atoms causes only negligible broadening in the specific
heat transition width compared to the pair disordered case. This
is shown in Fig. \ref{fig:Entropy}(b), where both curves are
produced from the impurity distribution in Fig.
\ref{fig:2Dmaps}(c). The transition is sharp because  potential
scatterers cause local suppressions of the order parameter,
whereas in interstitial regions $\Delta$ decreases with
temperature in a manner similar to the pure system. We have also
studied distributions of strong scatterers at the percent level,
and find similarly sharp transitions. On the other hand, potential
disorder smooth on a scale of $\xi_0$ can lead to large
modulations of $\mu$ which, even within BCS theory, will give
larger transition widths. In this case, however, the associated
local spectra are not consistent with STS measurements, as shown
in Ref. \onlinecite{NAMH05}.

As  evident from \ref{fig:Entropy}(a), pairing disorder with
parameters fixed to yield reasonable variations of the gap size
and graininess at low temperature\cite{NAMH05} leads to a
broadened transition width similar to the experimental
observations\cite{Loram}.
\begin{figure}[t]
\begin{minipage}{.99\columnwidth}
\includegraphics[width=.99\columnwidth,height=3.5cm]{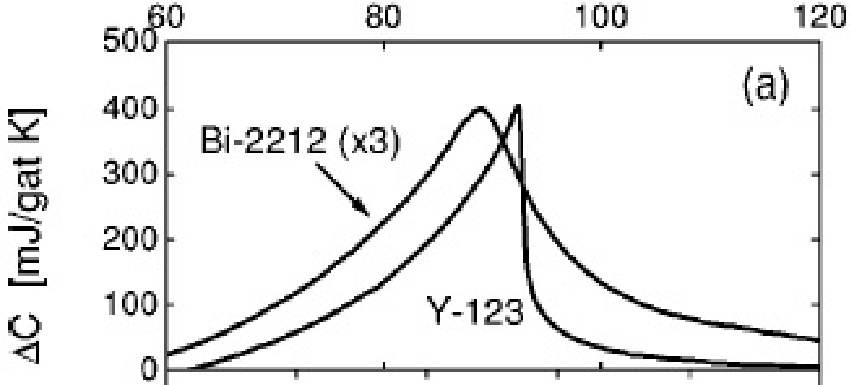}
\end{minipage}\\
\hspace{0.4cm}
\begin{minipage}{.94\columnwidth}
\includegraphics[width=.94\columnwidth,height=4.0cm]{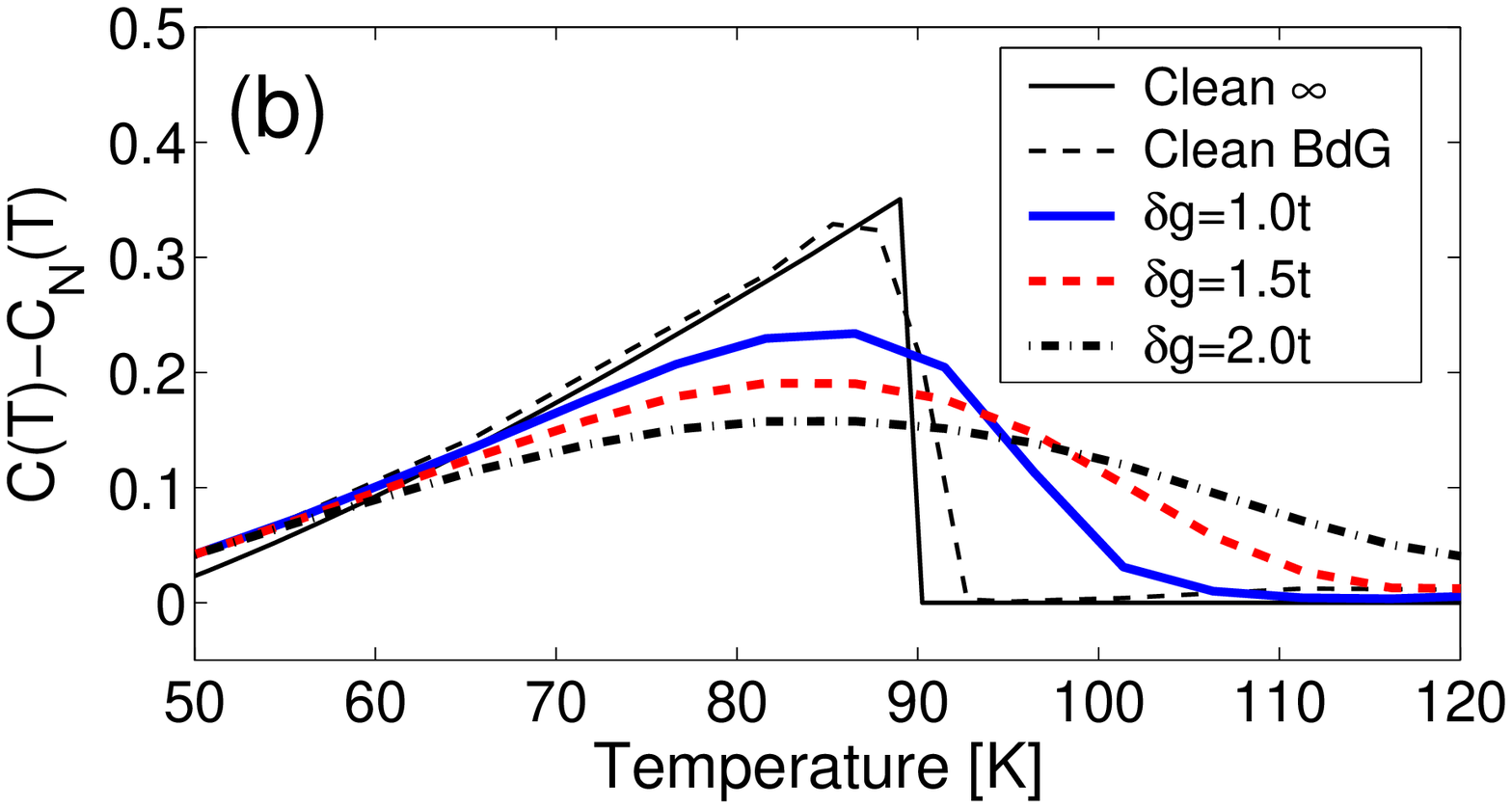}
\end{minipage}\\[-.1cm]
\caption{(Color online) (a) Specific heat transitions vs $T$ for
YBCO and BSCCO optimally doped powders, from Ref.
\onlinecite{Loram}. (b) $C(T)-C_N(T)$ ($N$: normal state) vs $T$
with parameters from Fig.\ref{fig:Entropy}(a), assuming that
$T_{c0}=0.182t$ equals $T_c=90$K.}\label{fig:Comp}
\end{figure}
This is shown more clearly in Fig. \ref{fig:Comp}, where we
compare the experimental specific heat $C(T)$ for clean YBCO and
optimally doped BSCCO near $T_c$\cite{Loram} with our results for
the clean and pair disordered case. It is clear that in the case
of YBCO, the transition is extremely narrow, and the deviations
from the mean-field treatment discussed here are consistent with
weak 3DXY critical fluctuations over a range of a few
degrees\cite{salamon3DXY,hardy3DXY}. On the other hand, in the
case of BSCCO, it has been clear for some time that the 3DXY
description of the specific heat transition is not appropriate,
despite the fact that according to a naive application of the
Ginzburg criterion 3DXY fluctuations should be more
visible\cite{junod99}, and it has in fact been discussed as closer
to a Bose-Einstein condensate (BEC) with specific heat exponent of
$\alpha=-1$, although measured values are closer to
$\alpha=-0.7$\cite{junod99}.
This interpretation remains controversial, however.  Other authors
have suggested that a 3DXY divergence in $C(T)$ cut off by bulk
nanoscale inhomogeneities over a small range $\Delta T_c/T_c \sim
0.03$ could be consistent with the data\cite{Schneider,Loram}.
Here we have put forward  a similar scenario, but our microscopic
model implies that the inhomogeneities  dominate over a larger
range $\Delta T_c/T_c \sim 0.1-0.2$ to be consistent with STS.
Additional contributions from critical fluctuations must be added
to the mean field effects discussed here to obtain a complete
description.

{\it Conclusions.} We have presented theoretical calculations for
$d$-wave superconductors with atomic scale pair disorder, using
impurity parameters appropriate to reproduce semi-quantitatively the
gap maps produced by STM experiments on optimally doped BSCCO, and
shown that the experimental specific heat transition in this system
can be explained by this model as well.  This suggests that
substantial nanoscale electronic inhomogeneity is characteristic of
the bulk BSCCO system.\\

We acknowledge valuable discussions with J. C. Davis, M. Gabay, N.
Goldenfeld, J. Loram, and Z. Te\v{s}anovi\'{c}, and thank J. C.
Davis and K. McElroy for sharing their data. Partial support for
this research (B. M. A. and P. J. H.) was provided by ONR
N00014-04-0060. T. S. N. was supported by the A. v. Humboldt
Foundation.


\begin{thebibliography}{99}
\bibitem{comment1} A small mismatch is introduced here because the
experimental FOV is slightly skewed, and will be close to, but not
exactly equal to, a $90\times 90$ system.
\bibitem{comment3} This value of the correlation is sensitive to the details of the impurity
potential: for instance, if one uses the raw LDOS data in Fig
\ref{fig:2Dmaps}(a) as input, it increases to 0.3.
\end{thebibliography}
\end{document}